\documentclass{ifacconf}

\usepackage{graphicx}      % include this line if your document contains figures
\usepackage{natbib}
\usepackage{amsmath} 
\usepackage{bbm}% required for bibliography
\usepackage{bbm}

\usepackage{algorithm}
\usepackage{algpseudocode}
\begin{document}
\begin{frontmatter}
\vspace{-20pt}

\makebox[0pt][c]{\raisebox{1pt}{%

  \parbox{\textwidth}{\centering

    \small © 2025. This work has been accepted to IFAC for publication under a Creative Commons Licence CC-BY-NC-ND and will be presented at the Modeling, Estimation and Control Conference (MECC 2025) in Pittsburgh, Pennsylvania, USA.}

}}
 
\vspace{-15pt}

\title{Adaptive Altitude Control of a Tethered Multirotor Autogyro under Varying Wind Speeds using Differential Rotor Braking}

\author[First]{Tasnia Noboni} 
\author[First]{Tuhin Das}

\address[First]{Department of Mech. and Aero. Eng., Univ. of Central Florida, FL 32816, USA (E-mail: Tasnia.Noboni@ucf.edu, Tuhin.Das@ucf.edu).}
%\address[Second]{Department of Mechanical and Aerospace Engineering, University of
%Central Florida, FL 32816, USA (e-mail: Tuhin.Das@ucf.edu)}

\begin{abstract}     % Abstract of 50--100 words
A tethered multirotor autogyro can function as an unmanned aerial vehicle for energy-efficient and prolonged deployment, as it uses the available wind energy to sustain flight. This article presents an adaptive altitude control strategy for such a device. At a constant wind speed, the equilibrium altitude can be approximated by a quadratic function of the pitch angle. The proposed adaptive control estimates the coefficients of this quadratic function. The estimates are used for altitude control and to attain the maximum altitude (and minimum horizontal drift) for a given wind speed. A feedback controller based on regenerative differential rotor braking is used as the actuation to modulate the autogyro's pitch angle. Implementation of the controller using a control-oriented, higher-order dynamic model demonstrates the controller’s capability to regulate the altitude and maintain stable flights under varying wind speeds. Based on the system's maximum altitude tracking performance, the adaptive control is adjusted to improve performance under substantial changes in wind speeds.
\end{abstract}

\begin{keyword}
Adaptive control, Tethered UAVs, Autorotation, Autogyro, Regenerative braking
\end{keyword}

\end{frontmatter}
%===============================================================================

\section{Introduction}
\vspace{-0.03in}
An autogyro is a rotorcraft that generates lift through autorotation of unpowered rotors in a sufficiently strong wind field. Such rotorcrafts, when tethered to the ground, hold the potential as an energy-efficient monitoring device due to their ability to use wind energy for long-duration deployment without relying on external power. Analyzing such tethered systems can provide valuable insights into the practical design of an efficient surveillance system.

Autogyro modeling using the blade element momentum (BEM) approach has evolved from assuming constant pitch rotor blades \citep{Glauert26} to incorporating linearly varying pitch with experimental validation \citep{wheatley1935aerodynamic}. It is different from conventional helicopter modeling as the latter assumes constant rotor speed, which is not valid for autogyros. In the literature, the generic rotorcraft model is extended to the autogyro configuration by introducing rotor speed degree of freedom \citep{lopez2004dynamics,thomson2005application}. Building upon the work in \cite{wheatley1935aerodynamic}, the steady-state behavior of autogyros and their feasibility for high-altitude power generation have been studied in \cite{mcconnell2022equilibrium}. Despite the potential of tethered autogyros in the surveillance sector, there are limited studies on the detailed dynamic modeling and control of such rotorcraft, with a few studies focusing on specific aspects of stability and control, \cite{noboni2025}. 

%Although in the literature, there are several control algorithms based on PID and linear quadratic regulators \citep{cast,dan} to address the challenges of autonomous untethered hovering of quadcopters. They are developed for simple hovering flight, lacking robustness and exhibiting poor tracking performance in the presence of modeling errors. A nonlinear adaptive tracking system for a quadrotor in the presence of modeling errors and disturbance uncertainties is described in \cite{islam2014nonlinear}. 

In the proposed autogyro system, the tether introduces an additional complexity to the flight control of quadrotors due to its coupling with translational and rotational dynamics of the system, influencing its maneuverability. The longitudinal stability of autogyros has been studied using a linearized model with a straight, massless tether \citep{rye1985longitudinal} and for configurations with a teetering rotor \citep{houston1998identification}. Our previous works \citep{10155811,noboni2024altitude} have demonstrated the viability of applying differential rotor braking regeneratively to control pitch and altitude using a reduced-order dynamic model of a tethered autogyro. It also shows that the system's equilibrium space is influenced by the pitch angle. The equilibrium altitude rises with the pitch angle up to a certain value before decreasing with a further increase in pitch angle. This equilibrium trend is also confirmed with a comprehensive dynamic model in \cite{noboni2025}, which relaxes the assumptions of average aerodynamic force and takes transient behavior into account. 

In this paper, we propose a nonlinear adaptive altitude control strategy \citep{slotine1991applied} for a quadcopter-based tethered autogyro using a comprehensive control-oriented dynamic model. Control actuation is based on regenerative differential rotor braking, which yields a net energy positive actuation for autorotating rotors. The control framework presented here focuses on the performance optimization of the system by maintaining an altitude close to the achievable maximum height in varying wind speeds. This also ensures minimization of horizontal drift of the rotorcraft from its point of deployment.

\section{Dynamic Model and Equilibria}
\subsection{System Description and Model Overview}
\label{model_sd}
\begin{figure}[htpb]
\begin{center}
\includegraphics[scale=0.36]{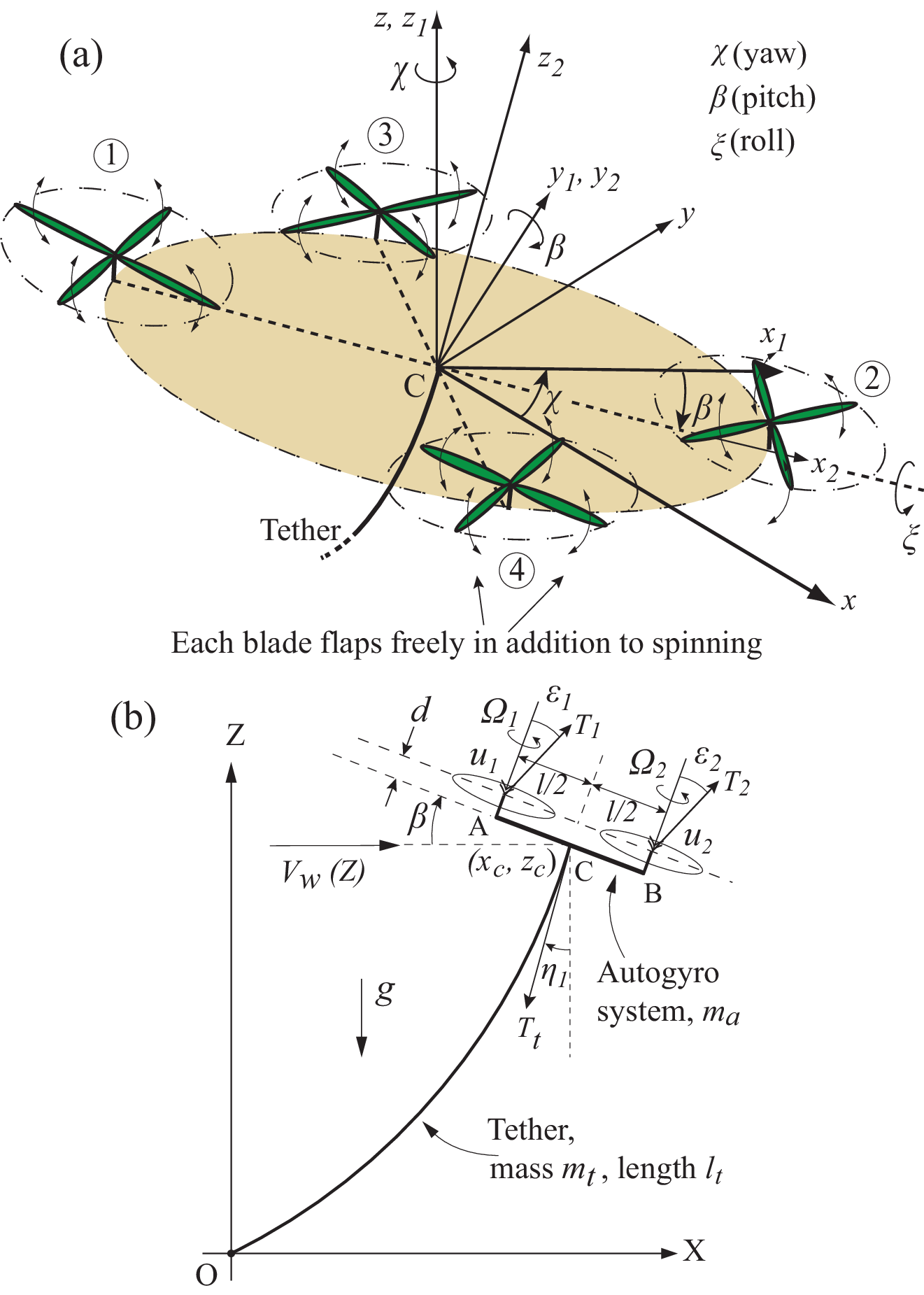}
	\caption{Tethered dual-rotor autogyro system: (a) Euler angles; (b) Dual rotor system in X-Z plane}
	\label{fig:sys}
\end{center}
\end{figure}
The autogyro studied in this paper has a quad-rotor mechanism with four equispaced blades per rotor, as shown in Fig.~\ref{fig:sys}(a). A two rotor version, constrained to the 2D X-Z plane is shown in Fig.~\ref{fig:sys}(b). The center of the frame $C$, located at $(x_c,\,z_c)$, is tethered to the ground. The frame has a pitch inclination of $\beta$. A hybrid dynamic model has been developed with the Lagrangian method, assuming a steady wind direction in the X-Z plane. 

\begin{figure} [tpb]
\begin{center}
\includegraphics[scale=0.34]{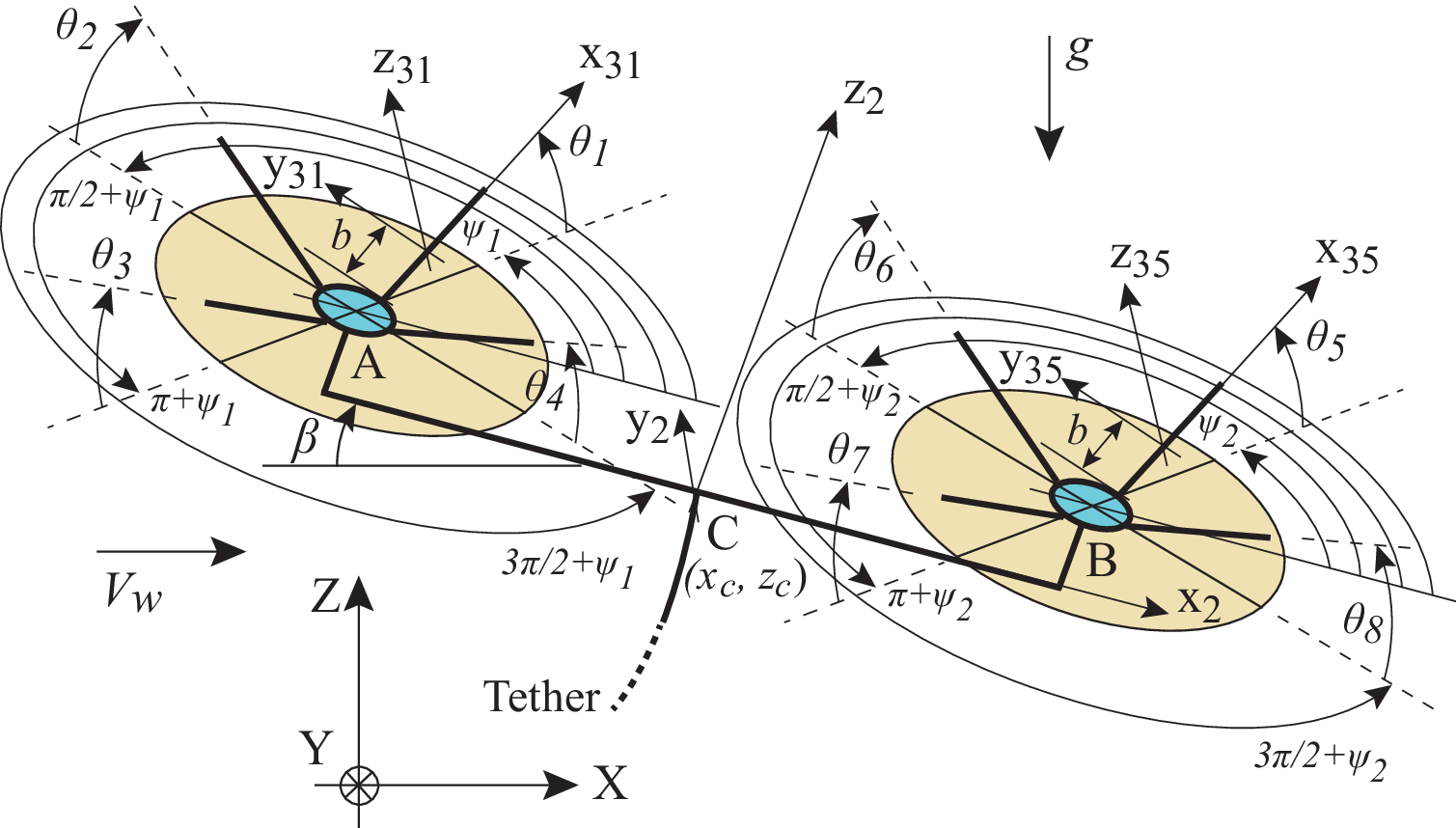}
	\caption{Reference frames and the generalized coordinates of the tethered autogyro in 2D}
	\label{fig:gc}
\end{center}
\end{figure}

The model is hybrid in the sense that the individual blade motion and aerodynamic forces are modeled in 3D, whereas the autogyro is constrained to the X-Z plane. Thereby, rotors 1 and 2 centered at $A$ and $B$ in Fig.~\ref{fig:sys}(b), have been modeled assuming that roll and yaw motions can be controlled by lateral rotors in the full 3D extension of the model. This setup leads to 13 generalized coordinates, illustrated in Fig.~\ref{fig:gc} and they are, 
\begin{equation}
    {q}=[x_c\; z_c\;  \beta\;  \psi_1\;  \theta_1\;  \theta_2\;  \theta_3\;  \theta_4\;  \psi_2\;  \theta_5\; \theta_6\; \theta_7\; \theta_8]^T
    \label{gen_coord}
\end{equation}
\noindent where, $\theta_j, \, j=1,2,\ldots,8$ is the flapping angle of each blade and $\psi_i,\,i=1,2$ indicate the rotational angles of the hubs in rotors 1 and 2, respectively. A $y-z-y$ Euler angle rotation sequence, see Fig.~\ref{fig:gc}, is used to obtain the orientation of each blade. The rotational sequence is defined by: 1) rotation by $\beta$ about $Y$ axis, 2) rotation by $(\psi_i+n\frac{\pi}{2})$ about $z_{2}$ direction, 3) rotation by $-\theta_j$ about $y_{3j}$, $j = 1,2,\dots 8$. Here, $i=1$, $n=(j-1)$ for $j=1,\ldots 4$ and $i=2$, $n=(j-5)$ for $j=5,\ldots 8$ following the convention used in Fig.~\ref{fig:gc}. 
The following equation is used to convert the coordinates from the inertial reference frame to the body-fixed reference frame of the blade,
\begin{equation}
    \left[ x_{3j}\,\, y_{3j}\,\, z_{3j} \right]^T = \mathbf{R_{-\theta_j,y}}\mathbf{R_{\psi_i,z}}\mathbf{R_{\beta,y}} \left[ X\,\, Y\,\, Z \right]^T
\end{equation}
where, $\mathbf{R_{-\theta_j,y}},\,\mathbf{R_{\psi_i,z}},$ and $\,\mathbf{R_{\beta,y}}$ are the rotation matrices about $y_{3j},\, z_2$ and $Y$ axes respectively shown in Fig.~\ref{fig:gc}. As the system is an assembly of the frame, two hubs at $A$ and $B$, and 8 blades, the kinetic energy $T$ and the potential energy $V$ terms are obtained for each component to generate the Lagrangian, i.e., $L=T-V$. %Differentiating $L$ with respect to $q_i$ and $\dot{q}_i$, the left-hand side of Eq.~\eqref{kinematic} is calculated. 
The equations of motion of the tethered system are,
\begin{equation}
    \frac{d}{dt}\left(\frac{\partial L}{\partial \dot{q}_i}\right) - \frac{\partial L}{\partial q_i} = Q_{gi}
    \label{kinematic}
\end{equation} 
where $i=1,\,2,\ldots 13$ and $q_i$ is the i\textsuperscript{th} generalized coordinate, see Eq.~\eqref{gen_coord}. Here, $Q_{gi}$ refers to the generalized forces and torques arising from aerodynamics and tether tension. Aerodynamic forces and moments are developed using the Blade Element Momentum theory \citep{Gessow52_b} using 10 discretized elements for each blade. Using static catenary mechanics \citep{Rimkus2013StabilityAO} and assuming that the tether is not subject to any aerodynamic loads, tether tension is modeled. To avoid any numerical instabilities caused by the taut tether, a compliance in the tether is introduced by adding stiffness \citep{masciola2013implementation}. Detailed expressions of generalized forces and torques are given in \cite{noboni2025}. Equation~\eqref{kinematic} is alternatively expressed as,
\begin{equation}
    \mathbf{A}\ddot{q}_i+\mathbf{B}=\mathbf{Q_{gi}} \quad \Rightarrow \quad \ddot{q}_i=\mathbf{A}^{-1}(\mathbf{Q_{gi}}-\mathbf{B})
    \label{kin_mat}
\end{equation} 
where, $\mathbf{A}$ is a $13 \times 13$ matrix dependent on $q_i$ and $\mathbf{B}$ is a $13 \times 1$ matrix dependent on both $q_i$ and $\dot{q}_i$. Equation~\eqref{kin_mat} is solved to obtain $\ddot{q}_i$. Successive numerical integration of $\ddot{q}_i$ yields $\dot{q}_i$ and ${q}_i$.

\subsection{Characteristics of Equilibria}
\label{EQB}
With suitable parameter values from \cite{10155811} given in Table \ref{TB:pres_tb} and initial guesses for states, the equations of motion in Eq.~\eqref{kin_mat} are solved.
\begin{table}[hbt!]
\begin{center}
\caption{Physical parameters of the system} \label{TB:pres_tb}
\begin{tabular}{cccc}
\hline
Parameter & Numerical Value (with units)\\  
\hline
\hline
$m_f$ & $13.6056~kg$\\
$m_h$ & $1~kg$ \\
$m_b$ & $2.5418~kg$ \\
$l$ & $8.13~m$  \\
$r_h$ & $0.0762~m$ \\
$d$&  $0.03048~m$\\
$R$ & $3.0480~m$ \\
\hline
\end{tabular}
\end{center}
\end{table}
A proportional controller for $\beta$ regulation ensures that the solution converges to equilibrium. The results provide insight into the equilibrium characteristics of the tethered autogyro, aiding the controller design studied in this paper. Definitions of all parameters are shown in Fig.~\ref{fig:sys}. The equilibrium characteristics of the tethered autogyro can be explained in the context of pitch angle, $\beta$, and the tip speed ratio $\mu$, i.e., the ratio of the wind speed parallel to the rotor disc to the speed of the rotor blade tip. The variable $\mu$ is calculated as $\mu = (V_w\cos{\beta}) / (\Omega R)$. Here, $\Omega$ is the rotor speed. 
\begin{figure}[htbp]
	\begin{center}
	\includegraphics[scale=0.092]{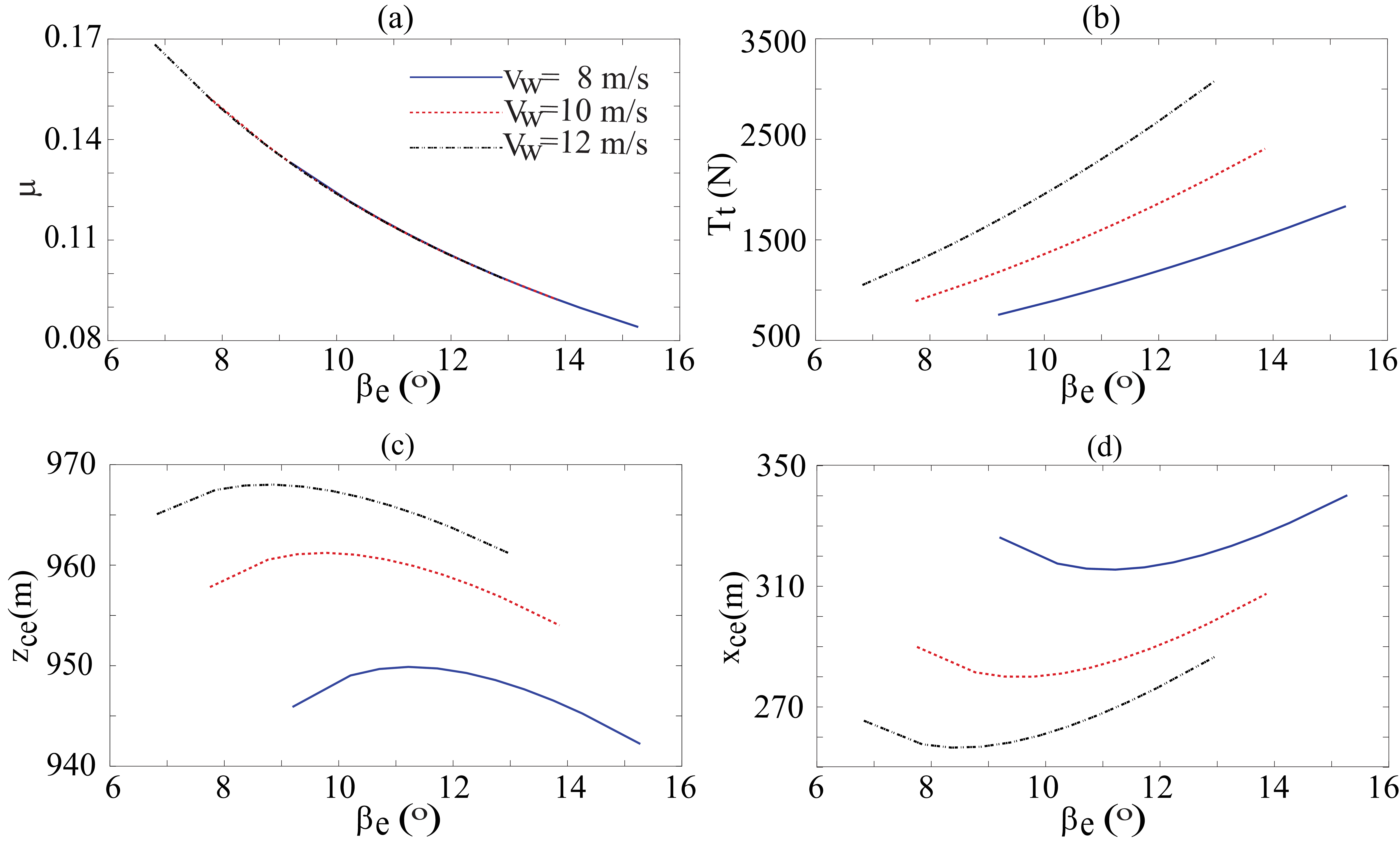}
            \vspace{-0.2in}
	    \caption{Equilibrium characteristics as functions of $\beta$ with ${l_t}$= 1000m: (a) Tip speed ratio; (b) Tether tension; (c) Altitude; (d) Lateral drift }
	\label{fig:beta_muTxz_vwvary}
    \end{center}
\end{figure}

Figure~\ref{fig:beta_muTxz_vwvary}(a) indicates that $\mu$ decreases with increasing $\beta$ and does not change if $\beta$ is kept constant in varying wind speeds. However, the range of $\mu$ gets wider with increasing $V_w$. Since momentum theory is valid approximately within $0.1<\mu<0.5$ \citep{wheatley1935aerodynamic,mcconnell2022equilibrium}, the system must operate within this range. The lower limit of $\mu$ corresponds to approximately $\beta =13^\circ$ in Fig.~\ref{fig:beta_muTxz_vwvary}(a). On the other hand, tether tension, $T_t$, becomes significantly high at the same $\beta$ as $V_w$ increases, shown in Fig.~\ref{fig:beta_muTxz_vwvary}(b), which suggests autogyro must shift to a lower $\beta$, if the tether tension needs to be regulated. 
%The variable $\mu$ used in Fig.~\ref{fig:beta_muTxz_vwvary}(a) refers to the tip speed ratio of either of the rotors, since in equilibrium, the tip speed ratios of both rotors have a negligible difference.
%\begin{figure}[htbp]
%	\begin{center}
	%\includegraphics[scale=0.12]{images/beta_xz_EQ.eps}
	%    \caption{Equilibrium characteristics as functions of $\beta$ with ${l_t}$= 1000~m: (a) altitude, (b) lateral drift}
       % \label{fig:beta_xz_vwvary}
    %\end{center}
%\vspace{-1in}
%\end{figure}

Equilibrium altitudes and lateral drifts of the autogyro also vary with pitch angle $\beta$ in different wind speeds as demonstrated in Figs.~\ref{fig:beta_muTxz_vwvary}(c) and~\ref{fig:beta_muTxz_vwvary}(d) respectively. With a fixed tether length and constant $V_w$, an increase in $\beta$ yields equilibrium altitude gain for the system up to a certain $\beta$. Further increase in $\beta$ results in altitude drop, increased drift, and progressively taut tether (see Fig.~\ref{fig:beta_muTxz_vwvary}(b)) as the drag force becomes dominant. Figure~\ref{fig:beta_muTxz_vwvary}(c) also suggests that with increasing $V_w$, the maximum altitude shifts toward lower $\beta$. However, the lower limit of $\beta$ is also restricted by larger lateral drift and low altitude due to the reduction of the lift force.

\subsection{Problem Definition and Control Approach}
\label{prob_st}
Equilibrium characteristics shown in Fig.~\ref{fig:beta_muTxz_vwvary} are crucial for formulating control solutions as the autogyro is expected to operate in an optimal range of $\mu$. The optimal range can be considered as one where higher elevation gain and lower lateral drift are achieved with a reasonable tether tension. In Fig.~\ref{fig:beta_muTxz_vwvary}, this corresponds to operating near the maximum altitude at different wind speeds. The one-to-one inverse relationship between $\mu$ and $\beta$, Fig.~\ref{fig:beta_muTxz_vwvary}(a), suggests that optimal $\mu$ can be maintained by modulating $\beta$. The optimal range, however, shifts with changing wind speeds, Fig.~\ref{fig:beta_muTxz_vwvary}(c). To operate close to the maximum altitude, in this paper we propose an adaptive adjustment of $\beta$ during flight, which is robust to wind speed fluctuations. 

In \cite{noboni2025}, the altitude control used a nested loop. The outer loop determined a reference $\beta$ based on altitude error while the inner loop determined the regenerative braking actuation based on the $\beta$ error. A limitation of this control is that for stability, the autogyro is required to operate on either the left or the right side exclusively of the maximum altitude point, Fig.~\ref{fig:beta_muTxz_vwvary}(c). To circumvent this issue, in this paper, we develop an adaptive strategy to estimate the $z$ vs. $\beta$ characteristics of Fig.~\ref{fig:beta_muTxz_vwvary}(c) using a polynomial approximation. This estimation drives the choice of the operating/optimal $\beta$ which is achieved by regenerative braking actuation.

%However, maintaining the optimal $\mu$ or $\beta$ is not accurately possible as $V_w$ is not constant in practical scenarios. Sudden change in $V_w$ can shift the optimal $\mu$ range, (see Fig.~\ref{fig:beta_muTxz_vwvary}(a)), causing the autogyro’s current operating $\mu$ to go beyond this range. The one-to-one inverse relationship between $\mu$ and $\beta$ suggests that optimal $\mu$ can be maintained by modulating $\beta$. The modulation of $\beta$ can be achieved in two ways. One approach is selecting the operating $\beta$ depending on the wind condition from a set of predefined values based on observed equilibrium characteristics. Another method requires adaptive adjustment of $\beta$ during the flight of the autogyro. Here, we take the latter approach of adjusting $\beta$ adaptively to control altitude.

\section{Control Design}
\label{CL}
%We investigate the altitude control problem in two steps. At first, we develop an adaptive controller for generating a pitch angle near a steady operating point by approximating the altitude as a function of pitch angle. The controller estimates the coefficients of this function while ensuring the system remains stable. In the second step, the estimated pitch angle is used as a reference input to a feedback controller. The feedback controller produces braking torques to modulate the pitch angle of the system, subsequently achieving altitude regulation.

\subsection{Adaptive Estimation Algorithm}
\label{sec_adapt_alg}
It is evident from Fig.~\ref{fig:beta_muTxz_vwvary}(c) that $z_c$ and $\beta$ do not have a monotonic relation. Close to the maximum altitude, $z_c$ can roughly be approximated as a quadratic function of $\beta$ as,
\begin{equation}
    z_c=-a\beta^2+b\beta+c\,; \quad a > 0, b > 0
    \label{z_app}
\end{equation}
However, the true values of the coefficients $a, b$, and $c$ are unknown. We assume,
\begin{equation}
    \hat{z}_c=-\hat{a}(t)\beta^2+\hat{b}(t)\beta+\hat{c}(t)
    \label{zhat_app}
\end{equation}
where, $\hat{z}_c$, $\hat{a}$, $\hat{b}$ and $\hat{c}$ are the estimates of $z_c$, $a$, $b$ and $c$ respectively. We assume $\beta$ and $z_c$ are measured. Thus,
\begin{equation} 
    e_{zh}=z_c-\hat{z}_c=-e_a\beta^2+e_b\beta+e_c
    \label{error}
\end{equation}
where, $e_{a}=a-\hat{a}(t),\, e_{b}=b-\hat{b}(t)$, and $e_{c}=c-\hat{c}(t)$.
Differentiating Eq.~\eqref{error} we get, 
\begin{equation} 
    \dot{e}_{zh}=-\dot{e}_a\beta^2-2e_a\beta\dot{\beta}+ \dot{e}_b\beta+ e_b\dot{\beta}+\dot{e}_c
    \label{ez_dot}
\end{equation}
We impose that the dynamics $\dot{e}_{zh}=-ke_{zh}$ be achieved by proper estimation of the parameters $a$, $b$ and $c$. Note that these estimates can change as the wind speed changes. The terms in the right-hand side of Eq.~\eqref{ez_dot} can be written as,
\begin{subequations}
\begin{equation} 
    -\dot{e}_a\beta^2-2e_a\beta\dot{\beta}=-k_1e_{zh}
    \label{ez_alt1}
\end{equation}
\begin{equation} 
    \dot{e}_b\beta+e_b\dot{\beta}=-k_2e_{zh}
    \label{ez_alt2}
\end{equation}
\begin{equation} 
    \dot{e}_c=-k_3e_{zh}
    \label{ez_alt3}
\end{equation}
\end{subequations}
where, $k_1,\,k_2,\,k_3\,>0$ and $k=k_1+k_2+k_3$. Note that by assuming $a$, $b$ and $c$ to be slowly varying, $\dot{e}_a = -\dot{\hat a}$, $\dot{e}_b = -\dot{\hat b}$ and $\dot{e}_c = -\dot{\hat c}$. However, Eq.~\eqref{ez_alt1} and \eqref{ez_alt2} cannot be directly used as adaptation laws since $e_a$, $e_b$ and $e_c$ are unknown. Hence, we consider, 
\begin{equation} 
\begin{aligned}
    \dot{e}_a\beta^2=-\dot{\hat a}(t)\beta^2=k_1e_{zh}+f_a,\\
    \dot{e}_b\beta=-\dot{\hat b}(t)\beta=-k_2e_{zh}+f_b,\\
    \dot{e}_c=-\dot{\hat c}(t)=-k_3e_{zh}
    \end{aligned}
    \label{common2}
\end{equation}
where, $f_a$ and $f_b$ are unknown functions that will be determined. Substituting Eq.~\eqref{common2} into Eq.~\eqref{ez_dot} we get,
\begin{equation} 
 \begin{aligned}
    \dot{e}_{zh}&=-k_1e_{zh}-f_a-2e_a\beta\dot{\beta} -k_2e_{zh}+f_b+ e_b\dot{\beta}-k_3e_{zh}\\&=-ke_{zh}-(f_a+2e_a\beta\dot{\beta})+(f_b+e_b\dot{\beta})
     \end{aligned}
    \label{ez_dot_2}
\end{equation}
Next, we consider the Lyapunov function $V_{Lyap}=\frac{1}{2}e_{zh}^2$. Differentiating $V_{Lyap}$ with respect to time yields,
%\vspace{-0.1}
\begin{equation} 
    \dot{V}_{Lyap}=e_{zh}\dot{e}_{zh}=-ke_{zh}^2-(f_a+2e_a\beta\dot{\beta})e_{zh} +(f_b+e_b\dot{\beta})e_{zh}
    \label{lyap}
\end{equation}
Therefore, from Eq.~\eqref{lyap}, we choose the following expressions for $f_a$ and $f_b$,
\begin{equation} 
    f_a=sgn(e_{zh})2|e_{amax}||\beta||\dot{\beta}|, \,f_b=-sgn(e_{zh})|e_{bmax}||\dot{\beta}|
    \label{fa_fb}
\end{equation}
From Eqs.~\eqref{common2} and \eqref{fa_fb}, adaptation laws are derived as,
\begin{equation}
 \begin{gathered}
    \dot{\hat{a}}(t)=-\Big(k_1e_{zh}+sgn(e_{zh})2|e_{amax}||\beta||\dot{\beta}|\Big)/ \beta^2,\\
    \dot{\hat{b}}(t)=\Big(k_2e_{zh}+sgn(e_{zh})|e_{bmax}||\dot{\beta}|\Big)/\beta,\\
    \dot{\hat c}(t)=k_3e_{zh}
\end{gathered}
\label{adapt_law}
\end{equation}
%The terms $sgn(e_{zh})[|2e_{amax}||\beta||\dot{\beta}|]$ and $sgn(e_{zh})[|e_{bmax}||\dot{\beta}|]$ in Eq.~\eqref{adapt_law} have negligible effects on $\dot {\hat{a}}(t),\,\dot {\hat{b}}(t)$ and $\dot { \hat{c}}(t)$.
%Thus from Eq.~\eqref{common2}, adaptation laws are derived as,
%\begin{equation} 
   % \begin{aligned}
    %\dot{\hat a}(t)=-\Big(k_1e_{zh}\Big)/\beta^2;\\
    %\dot{\hat b}(t)\beta=\Big(k_2e_{zh}\Big)/\beta;\; \quad \dot{\hat c}(t)=k_3e_{zh}
    %\end{aligned}
   % \label{adapt_law2}
%\vspace{-0.2in}
%\end{equation}
The coefficients $\hat{a}(t),\,\hat{b}(t)$ and $\hat{c}(t)$ are calculated by integrating Eq.~\eqref{adapt_law}.
%\begin{equation}
    %\begin{gathered}
    %\hat{a}(t)=\int_{t_{k-1}}^{t_k} \ \dot{\hat{a}}(t)\, dt; \quad \hat{b}(t)=\int_{t_{k-1}}^{t_k} \ \dot{\hat{b}}(t)\, dt;\;\\ \hat{c}(t)=\int_{t_{k-1}}^{t_k} \ \dot{\hat{c}}(t)\, dt; 
    %\end{gathered}
%\label{abch}
%\end{equation}
The above choice of adaptation laws results in $\dot{V}_{Lyap} < -ke_{zh}^2 < 0$, ensuring $e_{zh} \to 0$. Note that a practical implementation will only have estimates of $e_{amax}$ and $e_{bmax}$. This will cause $e_{zh}$ to have a bounded behavior around $e_{zh} = 0$. The bound will be determined by the magnitude of $k$ and those of $e_{amax}$ and $e_{bmax}$. Increasing $k$ will ensure convergence of $e_{zh}$ within a sufficiently small envelope around $e_{zh} = 0$. The pitch angle $\beta$ that yields probable maximum altitude, i.e., the $\beta$ value of the vertex of the estimated $\hat {z}_c-\beta$ curve, is,
\begin{equation}
    \beta\vert_{zmax}={\hat{b}(t)}/{2\hat{a}(t)}
    \label{max_beta}
\end{equation}
It is noted here that since one equation is being used to adaptively estimate three parameters, the individual estimates may not be accurate. To alleviate this issue, in the adaptive estimation, we will impose the condition that $\beta\vert_{zmax}$ reduces with increase in wind speed, as evident from Fig.~\ref{fig:beta_muTxz_vwvary}(c). Algorithm~\ref{alg:algo2} is used to determine $\hat{a}(t),\,\hat{b}(t),\,\hat{c}(t)$ and $\beta\vert_{zmax}$ at each time step. 
%\begin{enumerate}
%	\item  Initial guesses for $\hat{a}(0),\,\hat{b}(0)$ and $\hat{a}(0)$ are provided to Eq.~\eqref{adapt_law} to perform integration and evaluate $\hat{a}(t),\,\hat{b}(t)$ and $\hat{c}(t)$.
%	\item Altitude, $\hat{z}_c$ is estimated using Eq.~\eqref{zhat_app} and compared with the current altitude, $z_c$.
%	\item The absolute error, $|e_{zh}|$, is calculated using Eq.~\eqref{error} to check whether the convergence criterion, $\epsilon_c<10^{-5}$, is met. If not, the new values of  $\hat{a}(t),\,\hat{b}(t)$ and $\hat{c}(t)$ are used as new initial guess.
%	\item Steps 1-3 are repeated until desired convergence is achieved for $\hat{z}_c$.
	%\item Once convergence is achieved for $\hat{z}_c$ at the current instant, steps 1-4 are repeated for the next time step using the current $\hat{a}(t),\,\hat{b}(t)$ and $\hat{a}(t)$ as the initial guesses for the next time step.
%\end{enumerate}

\begin{algorithm}[ht]
\caption{Estimation Algorithm}
\begin{algorithmic}[1]
    \State Initialize $\hat{a}(t)$, $\hat{b}(t)$, and $\hat{c}(t)$ with $\hat{a}(t-1)$, $\hat{b}(t-1)$, and $\hat{c}(t-1)$ respectively \& initialize $e_{zh}(t)=e_{zh}(t-1)$
    
    \While{$|e_{zh}|\geq 10^{-5}$}
        \State  Evaluate $\hat{a}(t),\,\hat{b}(t),\,\hat{c}(t)$ by integrating Eq.~\eqref{adapt_law}
        \State Calculate $\hat{z}_c$ using Eq.~\eqref{zhat_app} 
        \State Compute absolute error $|e_{zh}|$ using Eq.~\eqref{error} 
        \If{$|e_{zh}| < 10^{-5}$}
            \State Break 
        \Else
            \State Update $\hat{a}(t),\,\hat{b}(t),\,\hat{c}(t)$ using the current values as new initial guesses
        \EndIf
    \EndWhile
     \State Calculate $\hat{z}_c$ at current time step using Eq.~\eqref{zhat_app}
     \State Calculate $e_{zh}(t)$ at current time step using Eq.~\eqref{error}
     \State Calculate $\beta\vert_{zmax}$ at current time step using Eq.~\eqref{max_beta}
     %\State Use $\hat{a}(t)$, $\hat{b}(t)$, $\hat{c}(t)$ as the initial guesses for the next time step
\end{algorithmic}
\label{alg:algo2}
\end{algorithm}

\vspace{-0.1in}
\subsection{Control Loops and Control Design}
\vspace{-0.1in}
\label{sec_controller}
In this paper, altitude control of the tethered autogyro is achieved without actuating the tether length. 
\begin{figure}[hpbt]
%\vspace{-0.1in}
	\centering
	\includegraphics[scale=0.31]{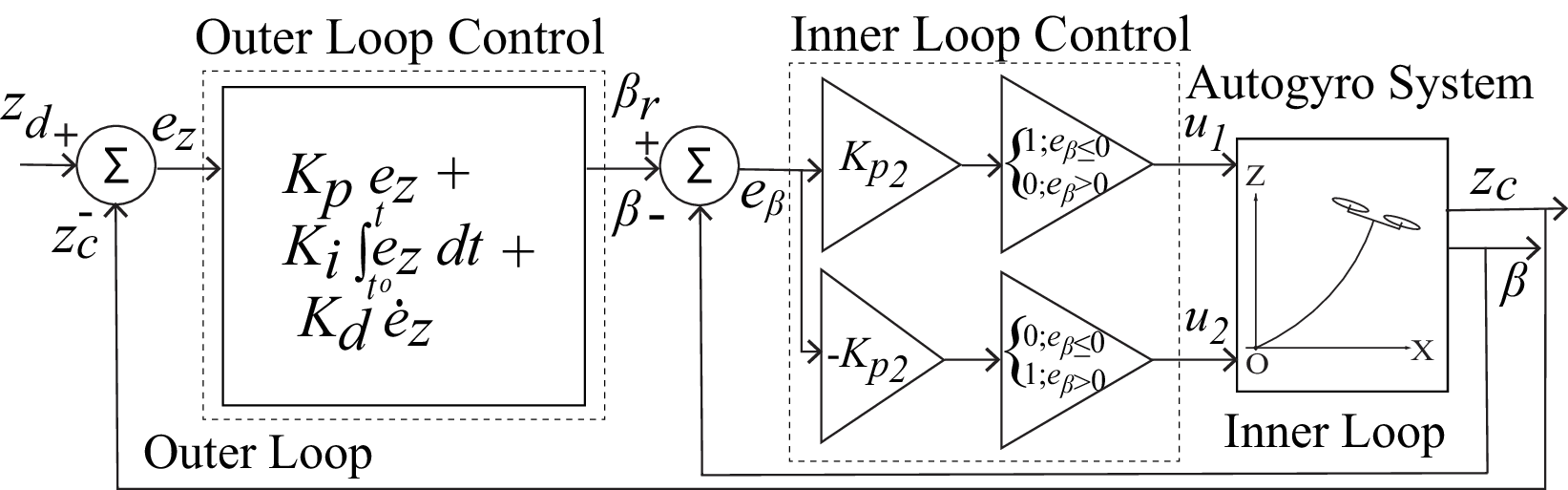}
    \vspace{-0.1in}
	\caption{Schematic of two-loop feedback controller \citep{noboni2025}}
       \label{fig:P_schm}
%\vspace{-0.1in}
\end{figure}
The control strategy utilizes a differential rotor braking method, where braking torques are applied in a regenerative manner to create a thrust imbalance between opposing rotors. The resulting thrust imbalance modulates $\beta$, which in turn leads to altitude change. These braking torques, denoted by $u_1$ and $u_2$, are chosen to be control inputs in this study. They are incorporated in the dynamics through equations of motion associated with the rotor speeds $\dot{\psi}_i$ in Eq.~\eqref{kinematic} as follows:
\begin{equation}
    \frac{d}{dt}\left(\frac{\partial L}{\partial \dot{\psi}_i}\right) - \frac{\partial L}{\partial \psi_i} = Q_{\psi_i}+u_i 
 \label{EOM_con}
\end{equation}
where, $i=1,2$. In Eq.~\eqref{EOM_con}, $u_1,u_2 \le 0$ ensures braking of the rotors. Altitude control is achieved by differential braking, i.e. $(u_1-u_2) \ne 0$, which alters $\beta$. 
\begin{figure}[hpbt]
	\centering
	\includegraphics[scale=0.38]{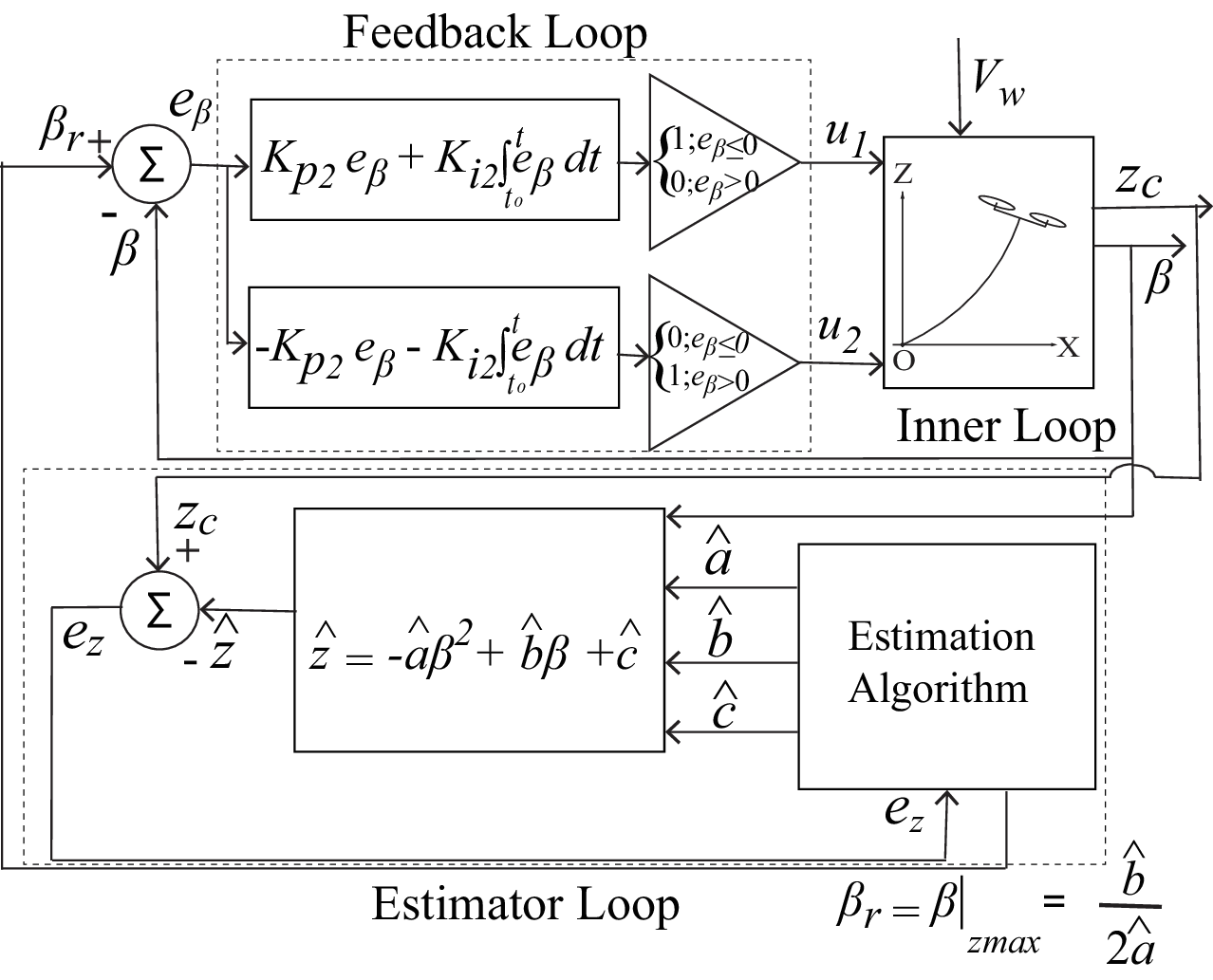}
    \vspace{-0.1in}
	\caption{Schematic of adaptive-feedback controller}
    \vspace{-0.1in}
    \label{fig:PI_schm}
\end{figure}
\begin{figure*}[tpb]
	\begin{center}
	\includegraphics[width=\textwidth]{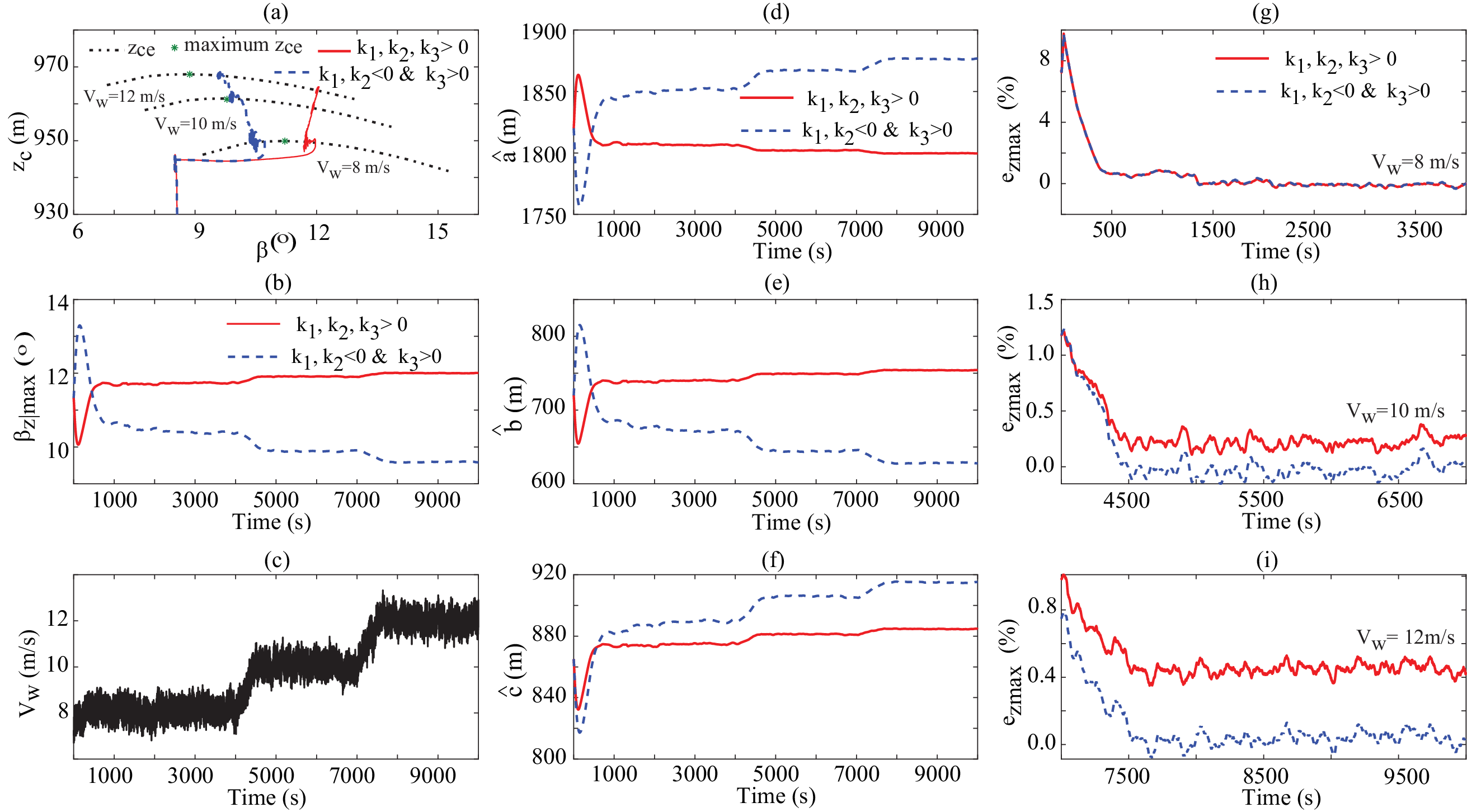}
    \vspace{-0.2in} 
    %\caption{Controller performance: (a) Variation of altitude with $\beta$; (b) Estimated $\beta\vert_{zmax}$; (c) Variable wind profile generated by TurbSim\citep{jonkman2014turbsim}; (d),(e),(f) Coefficients of estimated quadratic function $\hat{a},\,\hat{b},$ and $\hat{c}$ respectively; (g),(h),(i) Error between actual and estimated maximum altitude in $V_w$=8~m/s, 10~m/s, and 12~m/s respectively}
    \caption{Controller performance: (a) Variation of altitude with $\beta$; (b) Estimated $\beta\vert_{zmax}$; (c) Variable wind profile generated by TurbSim\citep{jonkman2014turbsim}; (d),(e),(f) Coefficients of estimated quadratic function $\hat{a},\,\hat{b},$ and $\hat{c}$ respectively; (g),(h),(i) Error between actual altitude and estimated maximum altitude in different average wind speeds}
    \label{fig:RS1}
    \end{center}   
\end{figure*}

Previously, we proposed a two-loop feedback controller \citep{noboni2025}, illustrated in Fig.~\ref{fig:P_schm}, to track a desired altitude. 
%emphasizing the robust stabilization of altitude without changing tether length.The employed control architecture is illustrated in Fig.~\ref{fig:P_schm}. 
A reference pitch angle, namely $\beta_r$, is generated based on the error between the desired and the actual altitude, i.e., $e_z= z_d-z_c$ by the outer loop. The inner loop generates either $u_1$ or $u_2$ depending on the sign of $e_\beta$. The resulting control law is,
\begin{equation}
\resizebox{\columnwidth}{!}{$
\begin{gathered}
\quad u_1=\left\{
\begin{array}{ll}
K_{p_2}(\beta_r -\beta) \\= K_{p_2}\left[\left(K_p e_z+K_i \int_{t_0}^{t} e_z \, dt+ K_d\dot{e}_z \right) -\beta \right];  & \quad \mbox{for} \quad \beta\ge \beta_r \\
0\,;& \quad \mbox{for} \quad \beta<\beta_r
\end{array}
\right. \\
\,\,\,\,\,\, u_2=\left\{
\begin{array}{ll}
0\, ;& \mbox{for} \quad \beta\ge \beta_r \\
-K_{p_2}(\beta_r -\beta) \\= -K_{p_2}\left[ \left( K_p e_z + K_i \int_{t_0}^{t} e_z \, dt+ K_d\dot{e}_z \right)-\beta \right]; & \mbox{for} \quad \beta<\beta_r
\end{array}
\right.
\end{gathered}
$}
\label{PID_control}
\end{equation}
where $K_p$, $K_i$, $K_d$ and $K_{p_2}$ are the gains of the controller. Results in \cite{noboni2025} suggest that the controller is effective in enabling the autogyro to track desired altitudes, $z_d$, within its operating region. The values of $z_d$ were selected arbitrarily, ensuring that the new altitude is reachable and smaller than the maximum attainable altitude at the corresponding wind speed, based on Fig.~\ref{fig:beta_muTxz_vwvary}(c). Specifically, for stability the controller required the target equilibria to be to the left of $\beta\vert_{zmax}$. This issue is alleviated in this work, where we use the aforementioned adaptive algorithm, Section~\ref{sec_adapt_alg}, to generate the reference pitch angle. Figure~\ref{fig:PI_schm} illustrates the closed-loop control architecture. Algorithm~\ref{alg:algo2} is used to determine the $\beta_r$ adaptively under varying wind speeds. The feedback loop uses a PI controller to maintain the generated $\beta_r$, enabling the autogyro to hover at the altitude corresponding to this angle. The PI controller is shown below,
\small
\begin{equation}
\scalebox{1}{$
\begin{gathered}
\quad u_1=\left\{
\begin{array}{ll}
K_{p_2}(\beta_r -\beta)+K_{i_2} \int_{t_0}^{t} (\beta_r -\beta) \, dt; & \;\, \mbox{for} \; \beta\ge \beta_r \\
%=K_{p_2}\Big(\frac{\hat{b}(t)}{2\hat{a}(t)}-\beta\Big)+K_{i_2} \int_{t_0}^{t} \Big(\frac{\hat{b}(t)}{2\hat{a}(t)}-\beta\Big) \, dt & \quad \mbox{for} \quad \beta\ge \beta_r \\
0\,;  & \;\,\mbox{for} \; \beta<\beta_r
\end{array}
\right. \\
\,\,\,\,\,\,u_2=\left\{
\begin{array}{ll}
0\,; & \mbox{for} \; \beta\ge \beta_r \\
-K_{p_2}(\beta_r -\beta) -K_{i_2} \int_{t_0}^{t} (\beta_r -\beta) \, dt\,;  & \mbox{for} \; \beta<\beta_r\\%=K_{p_2}\Big(\frac{\hat{b}(t)}{2\hat{a}(t)}-\beta\Big)+K_{i_2} \int_{t_0}^{t} \Big(\frac{\hat{b}(t)}{2\hat{a}(t)}-\beta\Big) \, dt & \mbox{for} \quad \beta<\beta_r
\end{array}
\right.
\end{gathered}
$}
\label{PI_control}
\end{equation}
\normalsize
where, $K_{p_2}$ and $K_{i_2}$ are the proportional and integral gain of the controller. Equation~\eqref{PI_control} is formulated ensuring $u_1,u_2\le0$ so that the control inputs can only be applied for braking the rotors. The values of $u_1$ and $u_2$ are restricted to small values to avoid abrupt fluctuations in $\beta$, ensuring robustness performance.

\vspace{-0.1in}
\section{Simulation Results}
\vspace{-0.1in}
The performance of the proposed controller with a fixed tether length of $1000$m, $K_{p_2}= 100$ and $K_{i_2}= 0.5$ is demonstrated through simulations. In Fig.~\ref{fig:RS1}, the autogyro is allowed to settle at a steady operating point, i.e., at $\beta_r=8.5^{\circ}$, up to 1200 seconds before the estimator loop is closed for $\beta_r$ generation. Figure~\ref{fig:RS1}(a) demonstrates the variation of altitude with $\beta$ along with the equilibrium $z_{ce}$ and $\beta_e$ from Fig.~\ref{fig:beta_muTxz_vwvary}(c) in varying wind speeds. 
The error between the actual maximum altitude based on Fig.~\ref{fig:beta_muTxz_vwvary}(c) and the $\hat{z}_c$ vertex of estimated $\hat {z}_c-\beta$ curve, namely $e_{zmax}$, for average $V_w$ of 8~m/s approaches nearly 0, as shown in Fig.~\ref{fig:RS1}(g). However, substantial change in the average values of $V_w$, such as from 8m/s to 10m/s and from 10m/s to 12m/s, shown in Fig.~\ref{fig:RS1}(c), causes $e_{zmax}$ to increase when $k_1, k_2 >0$ as evident in Figs.~\ref{fig:RS1}(h) and~\ref{fig:RS1}(i). This behavior is better visualized in Fig.~\ref{fig:RS1}(a), where the autogyro keeps hovering at altitudes corresponding to higher $\beta$ as $V_w$ increases, marked by the solid line. %This contradicts the equilibrium trend observed in Fig.~\ref{fig:beta_muTxz_vwvary}(c).
\begin{figure*}[htbp]
	\centering
	\includegraphics[scale=0.245]{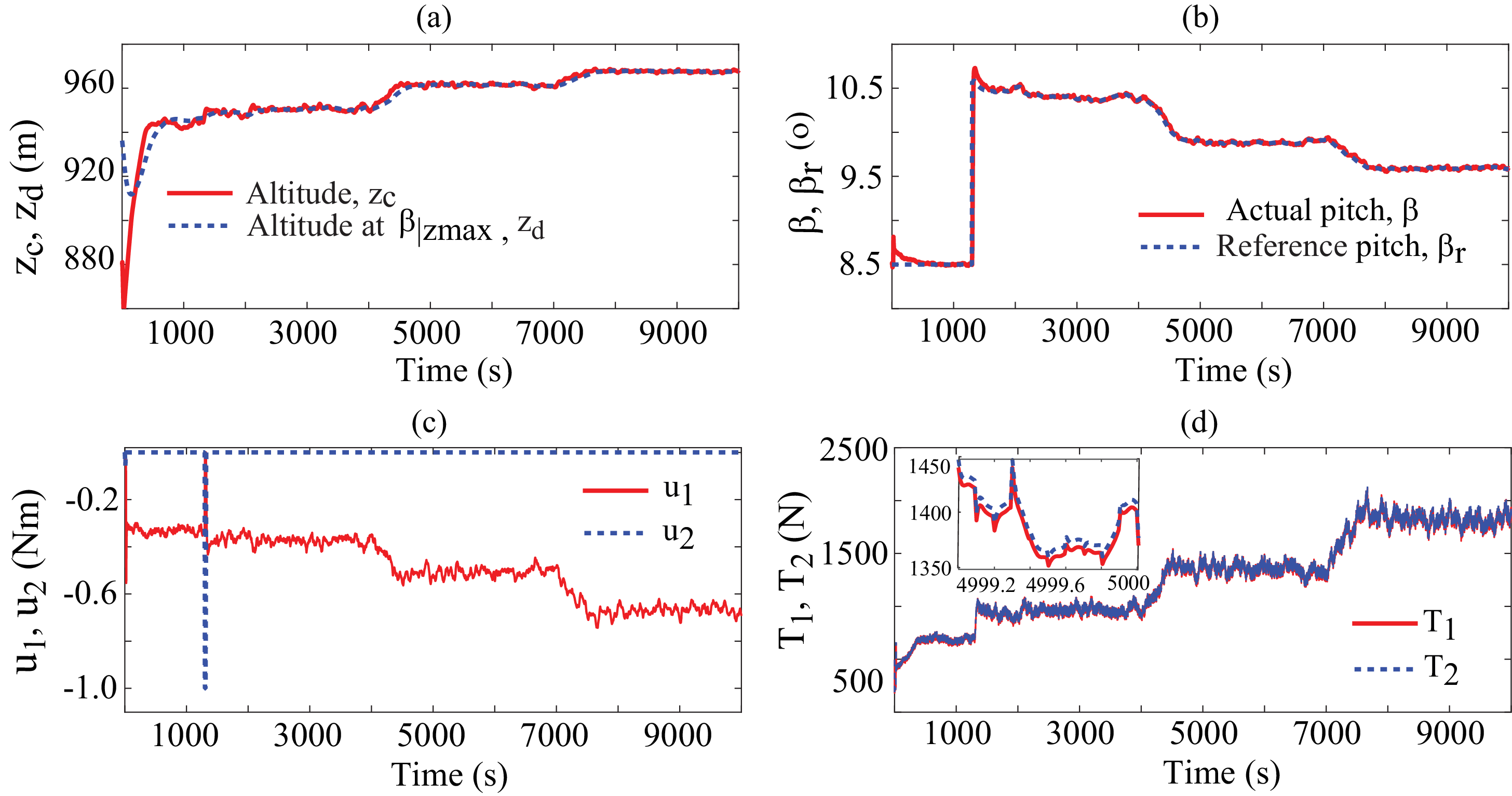}
        \vspace{-0.1in}
	    \caption{Controller performance: (a) Altitude; (b) Pitch angle; (c) Braking torques; (d) Thrust forces}
	\label{fig:Rs2_states}
\end{figure*}

Such a trend can be attributed to the simplifying assumptions made in developing the estimation algorithm in Section~\ref{sec_adapt_alg}. Only one equation, i.e. Eq.\eqref{z_app}, is used for estimating $\hat{a},\, \hat{b},\,$ and $\hat{c}$. The lack of information provided to the estimation algorithm results in the shifting of the extremum location away from the maximum altitude with increasing $V_w$. Equation~\eqref{max_beta} indicates that $\beta\vert_{zmax}$ is regulated by $\hat{a}$ and $\hat{b}$. Sudden increase in $V_w$ causes $z_c$ to increase, making $e_{zh} >0$ in Eq.~\eqref{error}. Positive $e_{zh}$ along with  $k_1, k_2 >0$ in Eq.~\eqref{adapt_law} yields a decrease in $\hat{a}$ and an increase in $\hat{b}$, thereby increasing $\beta\vert_{zmax}$ as observed in Figs.~\ref{fig:RS1}(d),~\ref{fig:RS1}(e) and~\ref{fig:RS1}(b) respectively. 

For shifting $\beta_{z|max}$ to the left with increasing $V_w$, $\hat{a}$ and $\hat{b}$ are expected to increase and decrease, respectively. Therefore, we propose $k_1<0$ and $k_2<0$ along with $k_3>0$ to achieve this goal. The adaptive gain $k_3$ is purposefully selected to be dominant over $k_1$ and $k_2$, i.e., $k_3> -(k_1 + k_2)$, so that $k$ in Eq.~\eqref{lyap} is always positive, thereby ensuring stability of the adaptive estimation algorithm. Simulation results exhibit that with $k_1$= -0.0003, $k_2$ = -0.003, and $k_3$= 0.01, $\beta\vert_{zmax}$ shifts toward left with increasing $V_w$, owing to the rising $\hat{a}$ and dropping $\hat{b}$, highlighted by dashed line in Figs.~\ref{fig:RS1}(a),~\ref{fig:RS1}(d), and~\ref{fig:RS1}(e). The coefficient $\hat{c}$ increases with $V_w$ for both cases as shown in Fig.~\ref{fig:RS1}(f). 
Figures~\ref{fig:RS1}(h) and~\ref{fig:RS1}(i) demonstrate that $e_{zmax}$ approaches to 0 for the average $V_w$ of 10~m/s and 12~m/s with $k_1,k_2<0$, thereby verifying the effectiveness of this adjustment.

Figure~\ref{fig:Rs2_states} demonstrates the closed-loop performance of the controller associated with changing $V_w$ (see, Fig.~\ref{fig:RS1}(c)) with $k_1,k_2<0$ and $k_3>0$. Figures~\ref {fig:Rs2_states}(a) and~\ref{fig:Rs2_states}(b) show that the system stabilizes at an altitude corresponding to $\beta_r=8.5^{\circ}$ using the PI controller. At 1200 seconds, the estimator loop is closed, which prompts $\beta_r$ to take the value of $\beta\vert_{zmax}$, and subsequently $z_c$ reaches the corresponding altitude. The integral action in the feedback loop eliminates the steady-state error between $\beta$ and $\beta_r$. The braking torques, in Fig.~\ref{fig:Rs2_states}(c), applied in each rotor are control inputs and restricted to be within $-1$~Nm$\le u_1,u_2\le0$~Nm to ensure a gradual change in $\beta$. Figure~\ref{fig:Rs2_states}(c) also corroborates that $u_1$ and $u_2$ are mutually exclusive as stated in Eq.~\eqref{PI_control}, i.e., when $u_1\neq0, u_2=0$ and vice versa. The zoomed view in Fig.~\ref{fig:Rs2_states}(d) illustrates that the thrust forces in the two rotors slightly vary from each other even when $z_c$ converges to the altitude corresponding to $\beta\vert_{zmax}$. This happens as rotor 1 requires continuous braking, shown in Fig.~\ref{fig:Rs2_states}(c), during the flight to maintain the desired pitch angle of $\beta\vert_{zmax}$.

\vspace{-0.05in}
\section{Conclusion}
\vspace{-0.1in}
An adaptive controller is developed for estimating the coefficients of the quadratic function that characterizes the equilibrium altitude-pitch angle relationship of a tethered autogyro. A feedback controller providing differential braking of the rotors utilizes these estimates for pitch modulation, thereby controlling the altitude. Simulation results show the efficacy of the proposed controller in controlling the altitude by maintaining a stable flight in a varying wind field. Additionally, modifications to the adaptive laws have been introduced to enhance the controller’s altitude tracking performance under significant wind speed variations. Future efforts will focus on refining the control algorithm by incorporating slope ($dz_c/d\beta$)-based error into the adaptation laws for more robust control analysis.

%As this work represents an initial step toward performance optimization of such systems, future efforts will focus on refining the control algorithm by incorporating slope ($dz_c/d\beta$) based error into the adaptation laws. Such an addition will assist in a more robust control analysis, facilitating practical deployment of tethered autogyros for surveillance applications.

%$\begin{ack}
%$Place acknowledgments here.
%\end{ack}

\vspace{-0.05in}
\bibliography{citation_file}             % bib file to produce the bibliography
                                                     % with bibtex (preferred)
                                                   
%\begin{thebibliography}{xx}  % you can also add the bibliography by hand

%\bibitem[Able(1956)]{Abl:56}
%B.C. Able.
%\newblock Nucleic acid content of microscope.
%\newblock \emph{Nature}, 135:\penalty0 7--9, 1956.

%\bibitem[Able et~al.(1954)Able, Tagg, and Rush]{AbTaRu:54}
%B.C. Able, R.A. Tagg, and M.~Rush.
%\newblock Enzyme-catalyzed cellular transanimations.
%\newblock In A.F. Round, editor, \emph{Advances in Enzymology}, volume~2, pages
%  125--247. Academic Press, New York, 3rd edition, 1954.

%\bibitem[Keohane(1958)]{Keo:58}
%R.~Keohane.
%\newblock \emph{Power and Interdependence: World Politics in Transitions}.
%\newblock Little, Brown \& Co., Boston, 1958.

%\bibitem[Powers(1985)]{Pow:85}
%T.~Powers.
%\newblock Is there a way out?
%\newblock \emph{Harpers}, pages 35--47, June 1985.

%\bibitem[Soukhanov(1992)]{Heritage:92}
%A.~H. Soukhanov, editor.
%\newblock \emph{{The American Heritage. Dictionary of the American Language}}.
%\newblock Houghton Mifflin Company, 1992.

%\end{thebibliography}

%\appendix
%\section{A summary of Latin grammar}    % Each appendix must have a short title.
%\section{Some Latin vocabulary}              % Sections and subsections are supported  
                                                                         % in the appendices.
\end{document}